\documentclass[twocolumn,showpacs,preprintnumbers,amsmath,amssymb]{revtex4}

\usepackage{graphicx}
\usepackage{dcolumn}
\usepackage{bm}

\begin{document}

\preprint{AJP}

\title{Advantages of adopting the operational definition of weight}

\author{Jos\'{e} M. L. Figueiredo}
 \email{jlongras@ualg.pt}
\affiliation{Departamento de F\'{\i}sica da Faculdade de Ci\^{e}ncias e Tecnologia da Universidade
do Algarve, Campus de Gambelas, 8005-139 FARO, Portugal
}%

\date{\today}

\begin{abstract}
With rare exceptions, in high school and college/university physics courses literature and in
journals of physics, the weight is defined as the gravitational force or an exclusive consequence
of it. These definitions lack logic from the perspective of present knowledge and can be
misleading. The operational definition of weight of a body as the force the body exerts on its
support or suspender can eliminate the ambiguities associated to ``true weight", ``apparent
weight", ``state of weightlessness", ``zero weight", ``zero gravity", ``microgravity", ``vertical",
``up" and ``down". Because the concept of weight of a body is not fundamental in Physics its
exclusion from the physics vocabulary will eliminate the need of some of former concepts or their
ambiguousness with evident benefit for physics teaching and learning. This paper discusses weight
of a body concepts and the advantages of adopting the above-mentioned operational definition. It is
believed this will eliminate the frequent identification of the body's weight with the body's mass
and will boost student's understanding of associated subjects.
\end{abstract}



\maketitle

\section{Introduction}

Everyone who teaches introductory physics is used to deal with the confusion students experience
when first confronted with the differences between concepts of mass and weight, and with the
ambiguities of the notions related to the weight definitions based on the gravitational force.
Several studies have shown students still have some misconception about the basic physics related
to weight and have difficulties applying them especially in imponderability or accelerated
environments \cite{Galili}\cite{GurelAcar}\cite{Sharma}. This is evident when they are asked to
explain what happens if everyday events such as walking were to take place in a ``weightlessness"
environments. The students often think ``microgravity", ``zero gravity", or ``weightlessness"
situations refer to events occurring ``outside" the Earth's or other celestial body's gravitational
influence, and they are surprised to hear that during a typical shuttle flight mission in an orbit
at an altitude of 400 km the gravitational force of the Earth is only 12\% less than at the Earth's
surface.

In the literature there are several definitions of weight based on the gravitational force: ``the
weight is the Earth gravitational force" \cite{Serway}; ``the force exerted by the Earth on an
object is called the weight of the object" \cite{Serway}; ``the weight of a body is the total
gravitational force exerted on the body by all other bodies in the universe" \cite{YoungFreedman}.
Frequently, the weight is considered a fundamental property of matter under the influence of a
gravitational field. Furthermore, in 1901 the Conf\'{e}rence G\'{e}n\'{e}rale des Poids et Mesures
declared ``the weight of a body is the product of its mass and the acceleration due to gravity"
\cite{3CGPM}.

These gravitational based definitions of weight are widely used in despite of the fact they are not
entirely satisfactory at the present knowledge and sometimes are even misleading. In addition,
there are a number of ambiguities associated with the weight gravitational definitions on the
meaning of weight and weight-related concepts such as ``true weight", ``apparent weight",
``weightlessness", ``zero-gravity", and ``vertical". That can easily lead to several misconceptions
which can contribute to widening the gap between what is taught and what is learned by the
students.

Very few authors adopt the alternative operational definition of weight of a body: ``the force
which a body exerts on its support or suspender that prevents its free fall" \cite{MarionHornyak}.
In this operational definition the weight is a force that results always from the direct contact of
the body with other body, i.e., the weight is a \textbf{contact} force.

This paper discusses weight, microgravity, weightlessness, vertical, up and down concepts, and the
advantages of the adoption of the operational definition of weight and/or the abandonment of weight
concept.

\section{Acceleration due to the gravity}

In the frame of the classical physics, the force of gravity is a long-range force, and, as far we
know, cannot be shielded \cite{gravitationalshielding}. In practical situations it is independent
of the state of motion of the objects. The acceleration due to the gravity corresponds to the
acceleration of the motion of a body as a result of the gravitational force, and in a given instant
equals the ratio of the gravitational force and the body amount of matter.

Accordingly the General Theory of Relativity the gravity corresponds to a modification (curvature)
in the space-time continuum caused by a concentration of mass or energy, that is, the space-time
geodesics surrounding substantial masses are curved lines and the bodies go through some form of
curved orbital path.

\subsection{Gravity under newtonian physics}

Since Isaac Newton presented the law of Universal Gravitation it is well accepted that the
gravitational interaction is universal and depends only on the body's quantity of matter and the
distance between their centers of mass. Following the works of Kepler and Galileu, Newton concluded
that the Earth's force of gravity \(\vec{F}_{g}\) exerted on our bodies or other mass \(m\) owing
to their gravitational interaction is given by
\begin{equation}\label{FgTm}
\vec{F}_{g}=-\frac{GMm}{|\vec{r}|^{3}}\vec{r}=\vec{\Gamma}m,
\end{equation}
where \(G\) is the gravitational constant (6.67\(\times10^{-11}\) N m\(^{2}\)kg\(^{-2}\)), \(M\) is
the Earth's mass, \(\vec{r}\) is the position vector of the body center of mass relatively to the
Earth center of mass. The gravitational force the Earth exerts on the body \(\vec{F}_{g}\) can be
written as the product of the body's mass and the local acceleration due to the gravity,
\(\vec{F}_{g}=m\vec{g}\); as mentioned previously, the vector
\(\vec{g}=-\frac{GM}{|\vec{r}|^{3}}\vec{r}\) corresponds to the body's acceleration due to the
gravitational field \(\vec{\Gamma}\), and its magnitude is approximately equal to 9,8 m s\(^{-2}\)
at sea level.

\begin{figure}[hbt]
\begin{center}
\includegraphics{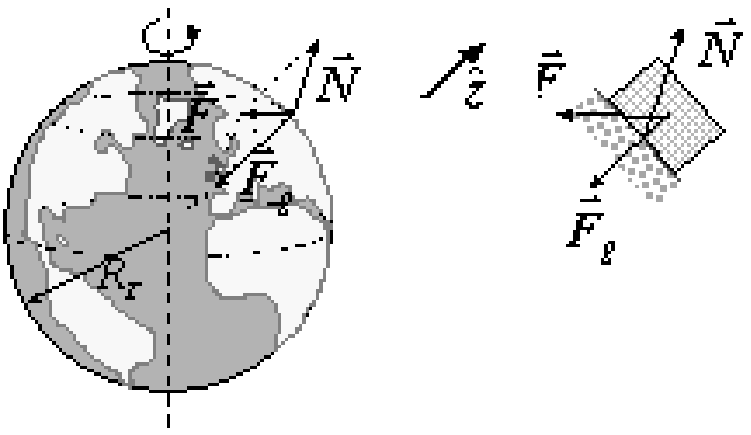}
\end{center}
\caption{\label{Fig1} Forces acting on a body at the the Earth's surface at a given latitude:
\(\vec{F}\) is the net force due to the force of gravity \(\vec{F}_{g}\) acting on the body and the
force of the reaction of the Earth's surface \(\vec{N}\) acting on the body.}
\end{figure}

The intensity of the force of gravity can be measured with the aid of a dynamometer or a spring
scale, provided that the body and the dynamometer are at rest relatively to the Earth. Let us
consider a body at rest on the surface of the Earth at a given latitude, Fig. \ref{Fig1}. The body
is acted upon by two forces: the force of gravity \(\vec{F}_{g}\) pointing towards the center of
the Earth and the force of the reaction of the Earth's surface (the support reaction force)
\(\vec{N}\), whose direction is determined not only by the force of gravity, but also by the
spinning of the Earth around its axis. Accordingly the second law of dynamics
(\(\vec{F}=d(m\vec{v})/dt\), where \(\vec{v}\) is the velocity of the mass \(m\)), the resultant
force \(\vec{F}\) of these two forces ensures the daily rotation of the body along the local
parallel. As a consequence the direction of the measured \(\vec{F}_{g}\) (and \(\vec{g}\)) differs
from the direction towards the center of the Earth - except at the poles and equator - by an angle
whose the maximum amplitude is less than 0.1\(^{0}\). In addition with the exception at the poles,
a scale or a dynamometer measures less than the gravitational force given by equation \ref{FgTm}
because a net force is needed to provide the centripetal acceleration needed to ensures the body
keeps up with the daily rotation of the Earth: the sensed acceleration of gravity is about 0.03 m
s\(^{-2}\) (0.35\% of \(g\)) less at the equator than at the poles, assuming a spherically
symmetric and homogenous Earth. Furthermore, the variation of density and the  surface
irregularities of the Earth give rise to a local changes in the gravitacional field and to the
vector \(\vec{g}\)).

Nevertheless, throughout the rest of the text we will consider the Earth as an homogenous sphere
and the effects of its rotation around its axis and the translation around the Sun or other motions
will be neglected because the values of the linear and the angular acceleration acquire by a body
due to these effect are very small when compared with the acceleration due to the gravity. For
simplicity, the Earth will be considered a frame of reference at rest during the characteristic
time of the phenomena analyzed here. The effect of the atmosphere will be also neglected. For
simplicity the gravitational influences of other celestial bodies is not considered.

\subsection{Gravity under the theory of general relativity}

The General Theory of Relativity addresses the problems of gravity and that of nonuniform or
accelerated motion. In one of his famous conceptual experiments Einstein concluded that it is not
possible to distinguish between a frame of reference at rest in a gravitational field and an
accelerated frame of reference in the absence of a significant gravitational field:
Einstein's principle of equivalence. From this principle of equivalence Einstein moved to a
geometric interpretation of gravitation: the presence of very large mass or a high concentration of
energy causes a local curvature in the space-time continuum. The space-time geodesic becomes curved
lines, that is, the space-time curvature is such that the \textbf{inertial }paths of a body is no
longer straight a line but some form of curved orbital path. Maintaining the classical view of teh
gravitation we can associate to the body's curved path motion a \emph{centripetal acceleration}
that is referred as the acceleration due to gravity.

\section{Operational definition of the force weight of a body}

What humans and matter experience as weight is not the force of gravity. What they experience as
weight is actually the consequence of the normal reaction of the ground (or whatever surface they
are in contact with or hang up) pushing upwards against them to counteract the force they are
exerting on the surface, the force weight of the body. A good evidence of this is given by the fact
that a person standing on a scale moving up and down on his toes does see the indicator moving,
telling that the measured force is changing while the gravity force, that depends only on the
person's and the Earth' masses and the distance between their centers of mass, does not vary to
induce such clear observable changes on the scale meter. Another evidence happens when going
towards the Earth surface in an elevator one experiences a greater strain in the legs and feet when
the elevator is stopping than when it is stationary or moving with constant velocity because the
floor is pushing up harder on the feet.

\subsection{Body at rest}

\begin{figure}[hbt]
\begin{center}
\includegraphics{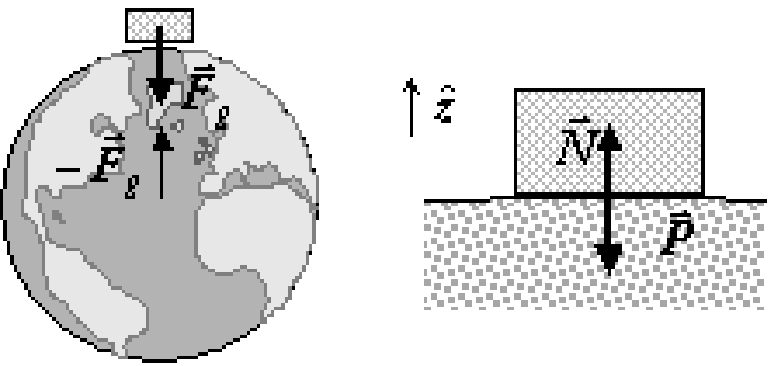}
\end{center}
\caption{\label{Fig2}a) \(\vec{F}_{g}\) and \(-\vec{F}_{g}\): the action-reaction force pair due to
the gravitational interaction between the body and the Earth. b) \(\vec{P}\) and \(\vec{N}\): the
action-reaction force pair due to the interaction between the body and the Earth surfaces.}
\end{figure}

Consider a body at rest on the surface of the Earth, Fig. \ref{Fig2}a. In this situation the body
experiences a force \(\vec{F}_{g}\) due to gravitational pull of the Earth. The reaction force to
this force is \(-\vec{F}_{g}\) and corresponds to the gravitational force exerted on the Earth by
the body. The force pair, \(\vec{F}_{g}\) and \(-\vec{F}_{g}\), consists on one force
\(\vec{F}_{g}\) that acts on the body and one force \(-\vec{F}_{g}\) that act on the Earth, and
constitutes an action-reaction pair.

The tendency of the body to accelerate towards the center of the Earth due to \(\vec{F}_{g}\) must
give rise to a force \(\vec{P}\), Fig. \ref{Fig2}b, force exerted by the body on the Earth surface.
If the body exerts on the Earth surface a force \(\vec{P}\), the Earth solid surface reacts
exerting a force \(\vec{N}\) on the body that balances the force \(\vec{P}\), Fig. \ref{Fig2}b. The
force \(\vec{N}\) is called the normal force and is the reaction to \(\vec{P}\):
\(\vec{N}=-\vec{P}\). The action the body exerts on the Earth (or other body) surface \(\vec{P}\)
corresponds to the operational definition of the force \textbf{weight of the body}.

Hence the body experiences no acceleration (it is at rest), the net force due to the two forces
acting on the body, \(\vec{F}_{g}\) towards the center of the Earth and \(\vec{N}\) outwards, is
null (Newton's second law of dynamics). Therefore, \(\vec{F}_{g}\) and \(\vec{N}\) are equal in
magnitude, opposite in orientation and have different application points. Similarly, and because
the Earth experiences no acceleration there are two equal and directly opposite forces acting on
the Earth, \(-\vec{F}_{g}\) applied on the Earth's center of mass and \(\vec{P}\) applied on the
Earth surface in contact with body. Although in this case \(|\vec{N}|=|\vec{F}_{g}|\) the normal
force \(\vec{N}\) is not the reaction to the gravitational force \(\vec{F}_{g}\) because this two
forces act on the body (as said previously \(\vec{N}\) is the reaction to \(\vec{P}\)).

Consider now the body is placed on (or hung on) a dynamometer-scale. When the body is placed on the
scale platform the dynamometer spring is compressed or extended (depending on the scale) and its
deformation is communicated to a calibrated dial read out. The body exerts an action \(\vec{P}\) on
the scale platform and through it on the spring. The scale dial reads the magnitude of the force
\(\vec{P}\) exerted \textbf{by} the body surface \textbf{on} the scale platform. By Newton's third
law of dynamics the scale platform reacts exerting a opposing force \(\vec{N}\) \textbf{on} the
body surface: both force have the same magnitude and directions but opposite orientations. It is
the force  \(\vec{N}\) that prevents the  body free fall towards the center of the Earth. The
weight is applied not to the body being considered itself, but to the scale platform.

If the scale is at rest relatively to the Earth then, as previously, the weight magnitude equals
the magnitude of the force of gravity acting on the body center of mass,
\(|\vec{P}|=|\vec{N}|=|\vec{F}_{g}|\). The weight and the force of gravity magnitudes are also
equal in the case of uniform and rectilinear motion of the scale and the body in a reference frame
associated with the Earth.

What happens when the scale and body are accelerating in relation to an frame of reference on the
Earth? This is the case, for example, of an elevator during stopping or starting. What is the scale
reading in these situations? During a sudden change in the elevator motion (on starting or braking,
for example) does remain valid the equality of the weight \(\vec{P}\) and gravity force
\(\vec{F}_{g}\) intensities?

\subsection{Body in a accelerated frame of reference}

Consider that a body of mass \(m\) is standing on a bathroom-type scale fixed in the floor of an
elevator with a TV camera circuit that will be used to record all the events. When the elevator
moves with an acceleration \(\vec{a}\), in accordance with Newton's second law of dynamics, as long
as the body and the scale surfaces are in contact the body moves together with the elevator and
scale with the acceleration \(\vec{a}\) under the action of two forces: the force of gravity
\(\vec{F}_{g}\) and the scale surface reaction force \(\vec{N}\) due to the body surface action on
the scale \(\vec{P}\). From Newton's second law of dynamics, \(\vec{F}=d(m\vec{v})/dt\), and
assuming the body's mass does not vary, the resultant of the forces acting on the body must equal
the product of its mass by its acceleration, which is the elevator acceleration
\(\vec{a}=d\vec{v}/dt\), that is,
\begin{equation}\label{000}
    m\vec{a}=\vec{F}_{g}+\vec{N}.
\end{equation}
Since the body's weight \(\vec{P}\) and the reaction force of the scale platform \(\vec{N}\)
constitutes a action-reaction pair, \(\vec{P}=-\vec{N}\), equation \ref{000} can be written as
\begin{equation}\label{00}
    \vec{P}=m\vec{g}-m\vec{a}.
\end{equation}

The magnitude of the body's action force (\emph{weight of the body}), \(|\vec{P}|\), is
proporcional to the value indicated in the scale.

Depending on the orientation of the acceleration \(\vec{a}=\pm a\vec{z}\) with \(a=|\vec{a}|\),
several situations may occur. However, here we discuss the cases of motion along the direction of
\(\vec{g}\), Fig. \ref{Fig2}.

Currently, the vertical is defined as the direction of plumb line that at the Earth surface and at
rest or on uniforme and rectilineal motion coincides with the direction of the gravity force.
However, a human being or other living being feels equilibrated in the direction of its weight
force. The concepts of vertical and down correspond to the direction and to the orientation of the
weight force, respectively. From equation \ref{00} one can conclude the vertical and the up and
down orientations depend essentially on the body state of motion characteristics, and contrary to
what is many times stated, the notions of vertical and up/down are not determined uniquely by the
gravitational force. The vertical is always the direction of the weight of the body and ``down"
corresponds to the weight force orientation. In accordance equation \ref{00}, to stay in
equilibrium during the bus starting movement we stoop forward and when it starts stopping we lean
backwards. In these situations our vertical is oblique and to not lose one's balance we align with
the new vertical defined by the direction of \(\vec{P}\).

In what follows, the axis \textit{oz} of the cartesian referential linked to the Earth coincides
with the direction along the center of the Earth and its positive orientation \(+\hat{z}\) points
outwards the center of the Earth.

\subsubsection{Elevator with uniform and rectilinear motion}

Since in this situation the body is not accelerating (\(\vec{a}=\vec{0}\)) the applied net force on
the body must be null. The equation \ref{00}, with \(\vec{g}=-g\hat{z}\), yields
\begin{equation}\label{333}
    \vec{P}=m\vec{g}=-mg\hat{z}.
\end{equation}

The intensity of the weight of a body standing on another body surface in uniform motion in a
straight line is the product of body's mass and the acceleration due to gravity, which coincides
with the definition of the 3rd CGPM \cite{3CGPM}.

\subsubsection{Elevator with acceleration opposite to the acceleration due to gravity}

\begin{figure}[hbt]
\begin{center}
\includegraphics{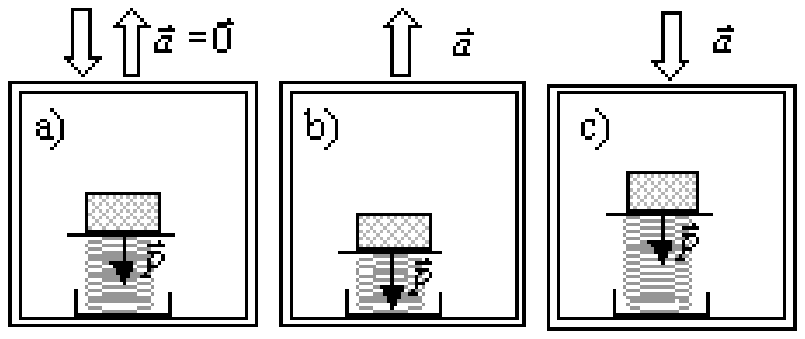}
\end{center}
\caption{\label{Fig3} Motion of a body siting on the platform of a spring-scale fixed to the
elevator floor: a) in uniforme motion, i.e., \(\vec{a}=\vec{0}\); b) moving against the gravity
force, i.e., \(\vec{a}=a\vec{z}\) with \(a>0\); c) moving moving with the gravity force, i.e.,
\(\vec{a}=-a\vec{z}\) with \(a>0\).}
\end{figure}

When the elevator is accelerating vertically with \(\vec{a}=a\hat{z}\), Fig. \ref{Fig3}b), the net
force acting on the body must be \(+m a\hat{z}\), and from equation \ref{00}
\begin{equation}\label{44}
        \vec{P}=-m(g+a)\hat{z},
\end{equation}
that is, the weight of the body intensity is greater than the product of its massa and the
acceleration due to gravity, i.e., \(P>mg\). The weight force maintains the orientation of the
acceleration due to gravity.

\subsubsection{Elevator with acceleration points towards the center of the Earth}

If the elevator is moving with an acceleration \(\vec{a}=-a\hat{z}\), Fig. \ref{Fig3}c), the net
force acting on the body is \(-m a\hat{z}\). There are three cases depending on the relative
magnitude of the elevator acceleration compared to the acceleration due to gravity \(g\): a)
\(0<a<g\); b) \(a=g\); c) \(a>g\).

\subsubsection*{a) \(0<a<g\)}

Because the elevator acceleration is inferior to the acceleration due to gravity the body keeps its
tendency to press the scale platform in order to move towards the center of the Earth with greater
acceleration. As a consequence the body presses the scale platform exerting a force \(\vec{P}\),
and the body moves together with the scale and the elevator with their acceleration
\(\vec{a}=-a\hat{z}\). In accordance with Newton's second law of dynamics, by equation \ref{00}
\begin{equation}\label{3-3}
    \vec{P}=-m(g-a)\hat{z},  ~ \textrm{with} ~ a>g,
\end{equation}
that is, the body weight intensity is smaller than the product of its mass and acceleration due to
gravity, \(P<mg\), but the weight force has the same orientation as the acceleration due to
gravity.

\subsubsection*{b) \(a=g\): free fall}

Lets consider the elevator free falling, that is, its acceleration is the acceleration due to
gravity \(\vec{a}\equiv\vec{g}=-g\hat{z}\). This happens when the elevator cables break. However,
it can be experienced without risk for few seconds in planes following parabolic trajectories or
much longer periods in spacecrafts orbiting the earth. All the bodies in the elevator are in free
fall moving with \(\vec{a}\equiv\vec{g}=-g\hat{z}\). From Newton's second law of dynamics, equation
\ref{00}, one gets
\begin{equation}\label{44}
         \vec{P}=\vec{0},
\end{equation}
that is, although the body does not move in relation to the scale, it exerts no action on the scale
surface. Therefore there is no reaction by the scale platform on the body and consequently the
scale dial shows zero weight.

The situation corresponds to the state of weightlessness, also known as imponderability. This state
is very often but misleading reported as zero-gravity state because the body is far from being away
of significant gravitational fields. The body is free falling together with the elevator and the
scale, and there are no forces pushing the body against the scale platform and \textit{vice-versa},
and not because the body and the elevator are outside the pull of the Earth's gravity. All objects
fall under the action of the Earth gravitational force (which is always present unless the Earth
``disappears").

\subsubsection*{c) \(a>g\)}

In accordance with equation \ref{00}, the weight of the body in an elevator with acceleration
\(a>g>0\) is given by:
\begin{equation}\label{33}
     \vec{P}=m(a-g)\hat{z}, ~ \textrm{with} ~ a>g,
\end{equation}
that is, when the elevator cabin and the scale move with an acceleration higher than the
acceleration of gravity the weight force would point against the acceleration due to gravity and
the weight intensity measured would be negative. The relation \ref{33} simply means the body loses
contact with the scale platform, which shows zero weight, unless is bounded to the scale. The body
is then in free fall and its acceleration \(\vec{a}\) is \(\vec{g}\). The scale and the elevator
move with acceleration \(a\) greater than \(g\) and as a consequence the body is left behind and
shortly is caught by the elevator ceiling.

Assuming there exist another (identical) scale fixed on the elevator ceiling facing down, and the
body rests on the ceiling scale platform, the weight force is
\begin{equation}\label{333}
     \vec{P}=m(a-g)\hat{z}, ~ \textrm{with} ~ a>g.
\end{equation}
Although the equations \ref{333} and \ref{33} are identical, the physical situations are distinct.
The intensity indicated by the second scale has physical meaning: the weight of the body is the
force the body exerts on the platform of the second scale, and its orientation is opposite to the
acceleration due to gravity. In contrast with the former situations, the orientation ``up" now
becomes ``down" and conversely.

\section{The weight force in a weak gravitational field}\label{p-verus-g}

The body weight force appears whenever the body's surface is constrained to interact directly with
the surface of another body. The weight of the body is opposite to the normal force (reaction
force) exerted by the surface where the body stands on or is in contact with, which prevents it
from moving through or away of the other body that is in contact with. The body action force or its
absence (weightlessness) does not depende of the existence of a gravitational field in the region
of the space where it is staying. Considere a spaceship in a region of the Universe where the
gravitational field is very small. The bodies in the interior of spaceship traveling in this region
with uniforme and rectilinear motion would experience zero weight, because they are in a
zero-gravity effective situation. Any spring-scale in contact and moving with them measures no
weight because the objects are not constrained to contact their surfaces to originate the normal
forces (zero action or zero weight gives rise to no normal force).

Lets now considere the spaceship turns on its engines. In the case of a spacecraft accelerating by
firing its rockets the thrust force is applied to the back end of the rocket by the gas escaping
out the back and the bodies in the interior of the veicule do not experiences weightlessness. The
rockets thrust force is transferred to each object in the spaceship through either pressure or
tension giving rise to the bodies action (weight force) on their supports or suspender. We can
conclude that the weight force in fact does not depend on the presence of a gravitational field.
Indeed, according the Einstein's Principle of Equivalence the bodies in a space veicule with an
acceleration \(\vec{a}\) in the absence of a gravitational field behave as the spaceship was at
rest or with constant velocity in a gravitational field with acceleration due to gravity
\(\vec{g}=-\vec{a}\). Taking in account the considerations made and the equation \ref{0000} the
weight of a body or the weightlessness state has nothing to do whether the body is under the
influence of a gravitational field or not. From equation \ref{00} results that if the spaceship is
accelerating uniformly out of the influence of a significant gravitational field, that is,
\(\vec{g}=\vec{0}\), the weight of a body carried by the vehicle is
\begin{equation}\label{0000}
    \vec{P}=-m\vec{a},
\end{equation}
that is, the weight force is opposite to the net force acting on the body and it is equal to the
product of the body's acceleration and mass, \(m\vec{a}\). In conclusion, the force the bodies
exert on their support (weight force) or their absence does not requires the presence of the
absence of a gravitational field. In the case of the presence of a gravitational field the force
the bodies exert on their supports depends also on characteristics of the relative motion between
the bodies and their supports.

As already mentioned the bodies in the interior of a spacecraft orbiting a celestial body, such as
the International Space station (ISS) around the Earth, are in a state of imponderability because
they do not exert any contact action on the other bodies. The weightlessness present several
challenges to the human organism which was designed to be live in a gravity environment and also
makes several of the mundane human actions, such as to walk, virtual impossible. Because in the
interior of the station there are no upward and downward convection currents of particles and gas
this has several effects on the human breathing system. The weightlessness also interferes with
cardiovascular system, with the heart beating faster because there's less resistance to the blood
flow. It is not possible to walk in weightlessness environment because the astronauts feet are not
constrained to the station pavement their feet action (weight force) on the surface of the station
is null. There is no normal reaction force and therefore the friction force is zero. It is the
friction force between the pavement and the astronauts feet that gives rise to the reaction force
needed to walk. This can also lead to the muscles atrophy, blood pump system malfunction and
difficult breathing.

Several plans have been proposed to create ``artificial gravity" in orbiting devices. The most
popular plan to produce ``artificial gravity" in vehicles designed to remain in orbit or stay in
out space for a long period of time are to set the spaceship into rotation with an angular velocity
\(\omega\) around its central axis. The bodies at any point at a distance \(r\) from the rotation
axis will experience a centripetal acceleration \(a=\omega^{2}r\). The weight of the bodies on the
outer rim of the spaceship opposes the centripetal force and its intensity is given by
\(P=m\omega^{2}r\).

\section{Imponderability and microgravity}\label{imp-micro}

In free fall all parts of an object accelerate uniformly and thus a human or other body would
experience no weight, assuming that there are no \emph{tidal forces} \footnote{The tidal force are
secondary effects of the forces of gravity due to Earth inhomogeneities and to the other celestial
objects gravity.}. The experience of no weight, by people and objects, is known as imponderability,
weightlessness or \emph{zero gravity}, although \emph{micro-gravity} is often used to describe such
a condition. Excluding spaceflight (orbital flight), weightlessness can be experienced only
briefly, around 30 seconds, as in an airplane following a ballistic parabolic path. In spaceships
the state of imponderabilidade or weightlessness can be experienced for extended periods of time if
the ship is outside the Earth's or other planet's atmosphere and as long as no propulsion is
applied and veicule is not rotating about its axis because the bodies in it interior are not
constrained to be in contact with other bodies or the station walls or floor. In particularly, the
astronauts are not pulled against the station pavement and, therefore, their bodies actions on the
surface of the station are null. In real free fall situations the tidal effects of the gravity on
the bodies, although small, are equivalente to a small acceleration  and the bodies are said to be
in a ``microgravity" environment because the weightlessness sensation is not complete.

The sate of imponderability experienced in orbiting spacecrafts is not as consequence of the small
value of the acceleration due to the gravity because the distance from the Earth.
Weightlessness is a consequence of the body and the spaceship accelerations to be only due to
gravity. The gravity acts directly on a person and other masses just like on the vehicle and the
person and the floor are not pushed toward each other. On the contrary, contact forces like
atmospheric drag and rocket thrust first act on the vehicle, and through the vehicle on the person.
As a consequence the person and the floor are pushed toward each other, giving rise to the weight
force.

As mentioned the term microgravity is usually used instead of weightlessness to refer the
environment within orbiting spacecraft. The use of the term micro-gravity without specifying its
exact meaning can strengthen the misconceptions associated to weight and gravitational force
because the term ``micro" could lead to the idea that acceleration due to gravity is very small
because the distance from Earth. To the contrary, the acceleration of the gravity due to the Earth
gravitational interaction is around 8.4 m s\(^{-2}\) at 400 km of height. Even it value at the
distance of the Moon orbit is 2.63\(\times10^{-3}\) m s\(^{-2}\), although in these regions the
acceleration due to Sun's gravity is near twice this value (\(\approx5.8\times10^{-3}\) m
s\(^{-2}\)). True Earth micro-gravity, \(g\simeq1\times10^{-6}\) m s\(^{-2}\), can be only
experienced at locations as far off as 17 times the Earth-Moon distance.

The term microgravity is more appropriate than ``zero weight" or ``zero-gravity" in the case of
orbiting spacecrafts because weightlessness is not perfect. Here the term microgravity does not
mean the acceleration due to gravity was strongly reduced but solely that its effects on the bodies
within the vehicle were substantially reduced. The term microgravity is used for the scientists to
characterize the residual acceleration experienced by the bodies in the interior of the spacecraft
as a consequence of forces between the bodies within the spaceship, the gravitational tidal forces
acting on the bodies and spacecraft and the atmosphere dragging force. These forces induce in the
bodies acceleration of intensities of some \(\mu\)m s\(^{-2}\), giving rise to the use of the term
``microgravity". For uncrewed spaceships free falling near the Earth it is possible to obtain 1
\(\mu g\); for crewed missions is difficult to achieve less than 100 \(\mu g\). The main reasons
are: i) the morphology of the Earth induce local gravitational variations; ii) the gravitational
effects of the other celestial bodies, especially the Moon and the Sun, which depend on their
relative position relatively to the Earth; iii) the acceleration due to gravity decreases one part
per million for every 3 m increase in height (in an orbiting spaceship the required centripetal
force and hence the acceleration due to gravity is higher at the far side than at the nearest side
of the ship relatively to the Earth); iv) although very thin, at for example 400 km of height, the
atmosphere gradually slows the spacecraft.

\section{Conclusion}

The identification of weight force as the force of gravity is misleading and lacks logic from the
perspective of the present knowledge. In the operational definition discussed the weight of a body
is the action force the body exerts on the surface of another body that it is in contact with, and
depends on their relative motion. Having in mind that the concept of weight is not fundamental in
Physics, the physics learning would benefit if the use of the vocable weight is avoided. One
advantage would be the rupture of the common sense identification between mass and weight force
concepts.
It is expected that this analysis will motivate physics instructors and authors, as well as the
scientific community, to replace the gravitational definition of weight by the operational one,
although its efective drooping is preferable.

\section*{Acknowledgments}

The author is grateful to Professor Robertus Potting, Dr. Paulo S\'{a}, Dr. Jos\'{e} Rodrigues, and
Dr. Alexandre Laugier for their comments and manuscript revision. The author acknowledges the
improvements that resulted from further discussions with other colleagues.

\end{document}


\preprint{GF RBEF}

\title[The concept of the weight: the operational definition as a contact force]
{É o conceito de ``peso do corpo" necessário em Física?}




\author{Jos\'{e} M. L. Figueiredo\footnote{E-mail: jlongras@ualg.pt}}
\address{Departamento de F\'{\i}sica da Faculdade de Ci\^{e}ncias e Tecnologia da Universidade do Algarve, Campus de Gambelas, 8005-139 FARO, Portugal}

\begin{abstract}

Com raras excepções, nos manuais de Física para os ensinos secundário e universitário e em muitas
publicações científicas, a força ``peso de um corpo" é definida como a força da gravidade ou uma
consequência exclusiva desta. 
Estas definições carecem de lógica do ponto de vista do conhecimento actual e podem ser mesmo
enganadoras para o estudante.  A definição operacional da força ``peso do corpo" como a força que o
corpo exerce no seu suporte tem a vantagem de eliminar a necessidade de vários conceitos associados
à definição gravitacional ou, pelo menos, a ambiguidade de alguns deles. Neste texto discute-se o
conceito da força ``peso" e apresentam-se exemplos das vantagens da sua redefinição e, em
particular, do abandono da expressão ``peso do corpo" uma vez que a noção de força ``peso do corpo"
não é fundamental em Física.


\end{abstract}
\date{\today}


\maketitle

\section{Introdução}

Quem lecciona disciplinas introdutórias de Física está acostumado a lidar com a confusão que os
estudantes fazem quando confrontados com os conceitos de massa e de força ``peso", e com as
ambiguidades das noções associadas à definição da força ``peso do corpo" baseada na força da
gravidade. Vários estudos indicam que os estudantes do secundário e da universidade mostram
dificuldades em aplicar os conceitos relacionados com a força ``peso" em situações de
imponderabilidade  e em sistemas com movimento acelerado
\cite{Galili}\cite{GurelAcar}\cite{Sharma}.
Com frequência, os alunos consideram que as situações correntemente designadas como
``gravidade-zero", ``microgravidade" ou ``sem peso" se referem a eventos que ocorrem fora da
influência gravitacional da Terra ou de outro corpo celeste e ficam, muitas vezes, surpreendidos
quando percebem que durante uma missão do vaivém espacial, por exemplo, a aceleração devida à
gravidade a que ficam sujeitos os astronautas é apenas 12\% inferior à que experimentam à
superfície da Terra.


Na literatura encontram-se várias definições da força ``peso" de um corpo baseadas na força da
gravidade: ``o peso" é a força gravitacional da Terra", ``a força exercida pela Terra num objecto é
chamada ``peso" do objecto" \cite{Serway}; ``o peso" de um corpo é a força gravitacional total
exercida no corpo por todos os outros corpos do Universo" \cite{YoungFreedman}. Estas definições
são compatíveis com a decisão da terceira reunião da Conferência Geral de Pesos e Medidas (CGPM),
em 1901, que definiu ``peso de um corpo" como o produto da sua massa pela aceleração devida à
gravidade \cite{3CGPM}.

As definições da força ``peso" baseadas na gravidade e os conceitos com ela relacionados são
largamente usadas nas aulas e nos livros de texto de Física, especialmente nas escolas secundárias
e nas disciplinas introdutórias de Física do ensino superior, apesar de serem insatisfatórias e
ambíguas, tendo em conta o conhecimento actual, e são, muitas vezes, origem de confusão.
%
Note-se, por exemplo, a imprecisão e a falta de clareza das noções de ``peso verdadeiro", de ``peso
aparente", de ``sem peso", de ``gravidade zero" e de ``microgravidade".
A ambiguidade dá origem a equívocos que podem facilmente contribuir para o alargamento do fosso
entre o que se pretende ensinar e o que é aprendido pelos estudantes.

Poucos autores adoptam a definição operacional da força ``peso de um corpo": ``força que o corpo
exerce no suporte a que está ligado" \cite{MarionHornyak}. Nesta definição operacional, a força
``peso do corpo" é uma força de ligação tal como, por exemplo, a força de atrito ou a tensão numa
corda. Este artigo discute o conceito de força ``peso do corpo", das noções de gravidade-zero, de
microgravidade, de vertical e de ``para baixo/cima", e as vantagens pedagógicas da adopção da
definição operacional da força ``peso" e/ou o abandono do termo força ``peso" em Física.


\section{Aceleração devida à gravidade}

No modelo da Física Clássica a força da gravidade é uma força de longo alcance entre os corpos com
massa e, tanto quanto se sabe, não pode ser blindada \cite{gravitationalshielding}. Em situações
convencionais é independente do estado de movimento dos corpos. A aceleração devida à gravidade é a
aceleração do movimento de um dado corpo material resultante apenas da acção da força da gravidade
sobre esse corpo e, num dado instante, corresponde à razão entre a força da gravidade e quantidade
de matéria do corpo.

Na descrição dada pela Teoria Geral da Relatividade um campo gravitacional corresponde a
modificações da geometria do espaço-tempo provocada por uma concentração de massa ou de energia. As
geodésicas do espaço-tempo são linhas curvas e os corpos descrevem órbitas correspondentes a estas
linhas curvas.


\subsection{Gravidade segundo a Física Newtoniana}

Desde que Isaac Newton propôs a lei da Gravitação Universal que se aceita que a interacção
gravitacional entre massas é universal e apenas depende das quantidades de matéria em interacção e
da distância que separa os seus centros de massa. A partir dos trabalhos de Kepler e de Galileu,
Newton concluiu que a força da gravidade \(\vec{F}_{g}\) que a Terra exerce num corpo de massa
\(m\), consequência da interacção gravitacional do corpo com o planeta Terra, é dada por
\begin{equation}\label{FgmM}
\vec{F}_{g}=-\frac{GMm}{|\vec{r}|^{3}}\vec{r}=\vec{G} m,
\end{equation}
onde \(G\) representa a constante gravitacional (6,67\(\times10^{-11}\) Nm\(^{2}\)kg\(^{-2}\)),
\(M\) é a massa da Terra, \(\vec{r}\) é o vector posição com origem no centro de massa da Terra e
extremidade no centro de massa do corpo.
A força da gravidade que a Terra exerce num corpo \(\vec{F}_{g}\) pode ser escrita como o produto
da massa \(m\) do corpo pela aceleração
devida à gravidade \(\vec{g}\), \(\vec{F}_{g}=m \vec{g}\); 
a grandeza \(\vec{g}=-\frac{GM}{|\vec{r}|^{3}}\vec{r}\) corresponde à aceleração que o corpo
adquire quando sujeito unicamente à acção do campo gravitacional \(\vec{G}\). Junto à superfície da
Terra \(\vec{g}\) é praticamente constante; a intensidade de \(\vec{g}\) ao nível do mar é,
aproximadamente, 9,8 m s\(^{-2}\).

\begin{figure}[hbt]
\begin{center}
\epsfbox{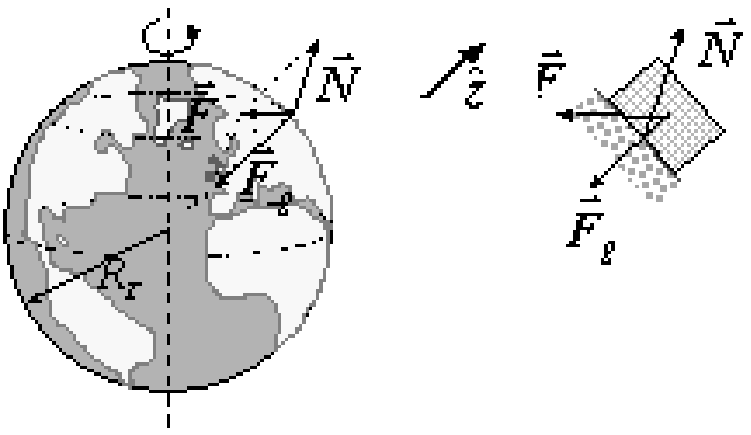}
\end{center}
\caption{\label{Fig1} Forças que actuam no corpo em repouso na superfície sólida da Terra, a uma
dada latitude: a força \(\vec{F}\) é a resultante da força da gravidade \(\vec{F}_{g}\) que actua
no corpo e da acção da superfície da Terra \(\vec{N}\) no corpo.}
\end{figure}

A intensidade da força da gravidade pode ser medida com auxílio de um dinamómetro ou de uma
balança-dinamómetro, assegurando que o corpo e o dinamómetro estão em repouso em relação à Terra.
Esta afirmação requer alguns esclarecimentos.
Considere-se um corpo em repouso relativamente à superfície da Terra a uma dada latitude, Fig.
\ref{Fig1}. O corpo é actuado por duas forças: a força da gravidade \(\vec{F}_{g}\), apontando para
o centro da Terra, e a força de reacção \(\vec{N}\) que a superfície da Terra (ou do suporte)
exerce sobre a superfície do corpo. A direcção de \(\vec{N}\) é determinada pela força da gravidade
e pela rotação da Terra em torno do seu eixo. Tendo presente a segunda lei de Newton da Dinâmica (a
resultante das forças que actuam num corpo é igual à taxa temporal de variação do momento linear do
corpo, \(\vec{F}=d(m\vec{v})/dt\)), a resultante \(\vec{F}\) destas duas forças assegura a rotação
diária do corpo segundo o paralelo que passa pela posição deste. Como consequência apenas deste
efeito, as direcções de \(\vec{F}_{g}\) e de \(\vec{g}\) ``medidas" diferem ligeiramente da
direcção do centro da Terra - excepto nos pólos e no equador - em um ângulo cuja amplitude é
inferior 0,1\(^{0}\). Verifica-se também que, devido à aceleração centrípeta do corpo, a
intensidade de \(\vec{F}_{g}\) indicada na balança ou no dinamómetro é inferior, excepto nos pólos,
ao valor dado pela equação \ref{FgmM}. Acresce ainda que as \emph{irregularidades} da superfície e
as variações de densidade nas diferentes regiões que constituem a Terra dão origem a um campo
gravitacional não exactamente central pelo menos nas proximidades da superfície da Terra e,
portanto, a variações na direcção e na intensidade de \(\vec{g}\).

Ao longo do resto do texto considera-se que a Terra é uma esfera homogénea e desprezam-se os
efeitos da rotação em torno do seu eixo e da translação em torno do Sol, e de quaisquer outros
movimentos, devido aos pequenos valores das acelerações linear e angular da Terra quando comparados
com aceleração devida à gravidade ou a outras forças aplicadas. Isto é, a Terra é considerada em
repouso durante os tempos característicos dos fenómenos aqui analisados. É desprezado também o
efeito da atmosfera no corpos.

\subsection{Gravidade segundo a Teoria da Relatividade Geral}

A Teoria da Relatividade Geral trata a gravidade e o movimento acelerado. Numa das suas famosas
experiências conceptuais, Einstein concluiu que não é possível distinguir entre um referencial em
repouso num campo gravitacional homogéneo e um referencial acelerado na ausência de um campo
gravitacional: princípio de equivalência de Einstein.
A partir deste princípio de equivalência, Einstein propõe uma interpretação geométrica da
gravitação: a presença de massa ou de energia provoca a curvatura do contínuo espaço-tempo, e as
geodésicas deixam de ser linhas rectas e passam a ser percursos curvos.
Numa descrição clássica deste efeito podemos associar ao movimento (órbita) de um corpo nas
proximidades da Terra, ou de outro corpo celeste, uma aceleração centrípeta que corresponderá ao
conceito clássico de \emph{aceleração devida à gravidade}.


\section{Definição operational de força ``peso do corpo"}


O que os seres humanos e a restante matéria experimentam como ``peso" não é a força da gravidade. A
sensação de ``peso" é devida à força normal que os suportes exercem nos corpos, constrangindo-os de
forma a contrariar a força (``peso") que eles exercem nos suportes. Uma evidência deste facto
ocorre quando uma pessoa em cima de uma balança-dinamómetro verifica que o valor indicado por esta
(a intensidade do ``peso") varia sempre que flecte as pernas sem perder o contacto com a superfície
da balança. Outra manifestação ocorre quando um elevador a descer, trava para parar: uma pessoa
sente um acréscimo de pressão nas pernas e nos pés. Estas variações não podem ser atribuídas à
força da gravidade, porque a distância entre os centros de massa da Terra e da pessoa praticamente
não se alterou, assim como as respectivas massas.

\subsection{Corpo em repouso}

\begin{figure}[hbt]
\begin{center}
\epsfbox{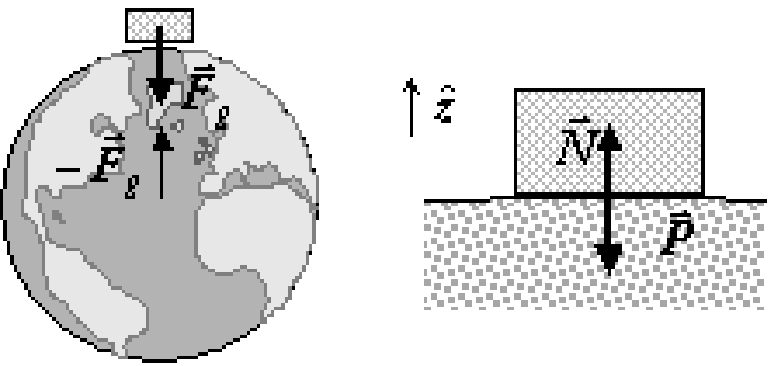}
\end{center}
\caption{\label{Fig2} a) \(\vec{F}_{g}\) e \(-\vec{F}_{g}\): par acção-reacção devido à interacção
gravitacional entre o corpo e a Terra. b) \(\vec{P}\) e \(\vec{N}\): par acção-reacção em resultado
da interacção entre as superfícies do corpo e da Terra.}
\end{figure}

Consideremos um corpo em repouso à superfície da Terra, Fig. \ref{Fig2}a. O corpo está sujeito à
força \(\vec{F}_{g}\) devido ao ``puxão" gravitacional da Terra. A força de reacção a esta força é
\(-\vec{F}_{g}\) e corresponde à acção gravitacional
exercida pelo corpo na Terra. O par de forças \(\vec{F}_{g}\) e \(-\vec{F}_{g}\) 
 constitui um par acção-reacção. A tendência do corpo em acelerar no sentido do centro da Terra devido à força \(\vec{F}_{g}\) dá
origem à força \(\vec{P}\), Fig. \ref{Fig2}b, força que a superfície do corpo exerce (acção do
corpo) na superfície da Terra.
Se o corpo exerce na superfície da Terra a força \(\vec{P}\), a superfície sólida da Terra reage
exercendo uma força \(\vec{N}\) no corpo que contrabalança a força
\(\vec{P}\), Fig. \ref{Fig2}b. A força \(\vec{N}\) 
designa-se força normal e é a reacção a \(\vec{P}\): \(\vec{P}\) e \(\vec{N}\) constituem um par
acção-reacção, \(\vec{N}=-\vec{P}\). A acção que o corpo exerce na superfície da Terra (ou suporte)
corresponde à definição operacional da força \textbf{``peso do corpo"}.

Em conclusão, no corpo actuam duas forças, a força \(\vec{F}_{g}\) apontando para o centro da Terra
e a força \(\vec{N}\) no sentido oposto. Uma vez que o corpo está em repouso relativamente à Terra
é nula a resultante das forças que actuam no corpo. Em consequência, as forças \(\vec{F}_{g}\) e
\(\vec{N}\) que actuam no corpo - em pontos de aplicação diferentes - têm a mesma intensidade e
orientações opostas. De forma similar, as duas força que actuam na Terra, \(-\vec{F}_{g}\) e
\(\vec{P}\), têm também a mesma magnitude, \(|\vec{P}|=|-\vec{F}_{g}|\), sentidos opostos e pontos
de aplicação diversos.

Consideremos agora o corpo pendurado num dinamómetro ou colocado no prato de uma
balança-dinamómetro. Quando o corpo é colocado na balança, a mola da balança é comprimida (no caso
do dinamómetro seria distendida) e a sua deformação é comunicada ao ponteiro da escala da balança.
O corpo exerce uma acção no prato da balança e através deste na mola. O ponteiro indica a magnitude
da força ``peso do corpo" \(\vec{P}\), força exercida \textbf{na} superfície do prato da balança
\textbf{pela} superfície do corpo.

Se o corpo e a balança estiverem em repouso em relação à Terra então, e como anteriormente, a
intensidade da força ``peso do corpo", que actua na balança, iguala a magnitude da força da
gravidade que actua no corpo, \(|\vec{P}|=|\vec{F}_{g}|\); a igualdade verifica-se também no caso
do corpo e da balança estarem em movimento uniforme e rectilíneo relativamente a um referencial
ligado à Terra.

Será que a relação \(|\vec{P}|=|\vec{F}_{g}|\) permanece válida se o corpo e o seu suporte
estiverem em movimento acelerado em relação à Terra? Qual é a indicação da balança nesta situação?

\subsection{Corpo num referêncial acelerado}

Imaginemos o corpo colocado no prato de uma balança fixa no pavimento de um elevador. Quando o
elevador acelera com uma aceleração \(\vec{a}\), enquanto estiver em contacto com a balança o corpo
move-se em conjunto com a balança e o elevador. No corpo actuam duas forças: a força da gravidade
\(\vec{F}_{g}\) e a reacção \(\vec{N}\) da balança à força ``peso do corpo" \(\vec{P}\). De acordo
com a segunda lei de Newton da Dinâmica e assumindo que a massa \(m\) do corpo não varia, a
resultante das forças que actuam no corpo é igual ao produto da massa \(m\) com a aceleração
\(\vec{a}\) do corpo, isto é,
\begin{equation}\label{000}
 m\vec{a}=\vec{F}_{g}+\vec{N}.
\end{equation}
Uma vez que a força ``peso do corpo" \(\vec{P}\) e a reacção \(\vec{N}\) da superfície da balança
formam um par acção-reacção, \(\vec{P}=-\vec{N}\), podemos rescrever a equação \ref{000} como
\begin{equation}\label{00}
  \vec{P}=m\vec{g}-m\vec{a}.
\end{equation}
Ter presente que o valor indicado na balança corresponde à intensidade da força \(\vec{P}\) que o
corpo exerce na balança.

Correntemente, define-se vertical de um lugar como a direcção da aceleração devida à gravidade.
Outras definições baseiam-se na direcção do fio-de-prumo, que à superfície da Terra e em repouso
coincide com a direcção da força da gravidade. A noção de vertical é mais geral. Por exemplo, um
ser humano ou outro ser vivo sente-se equilibrado na direcção da força que exerce no suporte e a
orientação ``para baixo" correspondem ao sentido dessa força. A vertical está sempre segundo a
linha de acção da força ``peso do corpo" e a orientação ``para baixo" corresponde ao sentido dessa
força. Da relação \ref{00} pode concluir-se que tanto a direcção da vertical de um corpo como as
orientações ``para baixo/cima" dependem da aceleração do corpo e da aceleração devida à gravidade.
Uma constatação quotidiana ocorre quando um autocarro arranca ou quando trava: nesta situação a
``nossa" vertical é oblíqua e para não nos desequilibrarmos inclinamo-nos na direcção da ``nova"
vertical.


No que se segue, consideram-se apenas as situações que envolvem movimento rectilíneo com aceleração
\(\vec{a}\) segundo uma linha de acção 0z que passa pelo centro da Terra e cujo sentido positivo,
\(+\hat{z}\), coincide com a orientação de afastamento do centro da Terra, Fig. \ref{Fig2}.

\subsubsection{Elevador com movimento rectilíneo e uniforme}

 Nesta situação a velocidade do corpo é constante, \(\vec{a}=\vec{0}\), Fig. \ref{Fig3}a, e
tendo em conta que \(\vec{g}=-g\hat{z}\) e \(g>0\), da equação \ref{00} resulta:
\begin{equation}\label{333}
    \vec{P}=-mg\hat{z}.
\end{equation}
Este resultado é idêntico ao obtido para um corpo em repouso à superfície da Terra. Neste caso, a
força peso do corpo coincide com o produto da sua massa pela aceleração devida à gravidade, em
acordo com a definição da terceira reunião da CGPM \cite{3CGPM}.

\subsubsection{Elevador com aceleração oposta à aceleração devida à gravidade}

\begin{figure}[hbt]
\begin{center}
\epsfbox{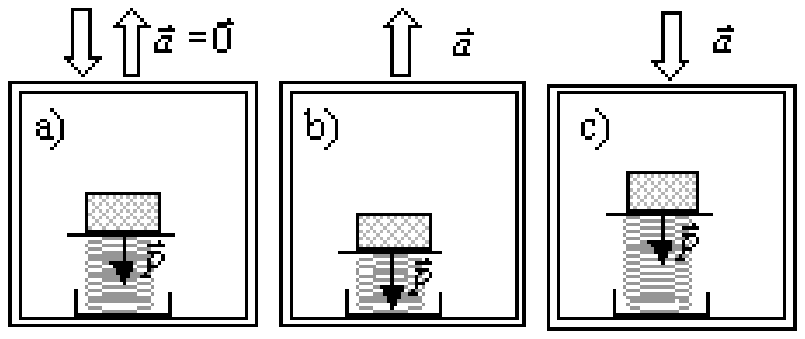}
\end{center}
\caption{\label{Fig3}Movimento do elevador contendo uma balança fixa ao pavimento, no prato da qual
é colocado o corpo em análise: a) movimento rectilíneo uniforme, \(\vec{a}=\vec{0}\); b) movimento
acelerado no sentido ascendente, \(\vec{a}=a\hat{z}\) e \(a>0\); c) movimento acelerado no sentido
descendente, \(\vec{a}=-a\hat{z}\) e \(a>0\).}
\end{figure}

Quando o elevador se afasta da superfície da Terra com aceleração \(\vec{a}=a\hat{z}\) e \(a>0\),
Fig. \ref{Fig3}b, a relação \ref{00} toma a forma:
\begin{equation}\label{44}
        \vec{P}=-m(g+a)\hat{z}.
\end{equation}
Neste caso, a magnitude da força ``peso do corpo" é superior a \(mg\). A força ``peso do corpo" tem
a orientação da aceleração devida à gravidade.

\subsubsection{Elevador com aceleração dirigida para o centro da Terra}


Se o elevador se deslocar com aceleração no sentido do centro da Terra - sentido da aceleração
devida à gravidade, \(\vec{a}=-a\hat{z}\) e
\(a>0\), Fig. \ref{Fig3}c, 
podem ocorrer três situações: \(a<g\), \(a=g\) e \(a>g\).

\subsubsection*{a) \textbf{\emph{a\(<\)g}}}

Como a intensidade \(a\) da aceleração do elevador é inferior à magnitude da aceleração devida à
gravidade \(g\), a tendência do corpo em se deslocar no sentido do centro da Terra com aceleração
\(\vec{g}=-g\hat{z}\) é contrariada pela oposição do prato da balança. Em resultado o corpo exerce
no prato a força \(\vec{P}\):  o corpo e o prato continuam ligados (em contacto), deslocando-se
ambos com a aceleração do elevador \(\vec{a}=-a\hat{z}\). Da relação \ref{00} obtém-se
\begin{equation}\label{3-3}
    \vec{P}=-m(g-a)\hat{z}, ~ \textrm{com} ~ 0<a<g.
\end{equation}

O ``peso" mantém a mesma orientação que a força da gravidade, embora com magnitude inferior a
\(mg\).

\subsubsection*{b) \(\textbf{\emph{a=g}}\): queda livre}\label{free-fall}


Quando o elevador está em queda livre, isto é, está apenas sujeito à acção da gravidade, todos os
corpos no seu interior estão também em queda livre e, portanto, movem-se todos com aceleração
\(\vec{a}=\vec{g}\), e, embora o corpo não se desloque em relação à balança, também não exerce
qualquer acção na superfície do prato. Neste caso, da equação \ref{00} resulta:
\begin{equation}\label{44}
         \vec{P}=\vec{0}.
\end{equation}

Esta situação corresponde ao estado de imponderabilidade e ocorre, por exemplo, quando os cabos do
elevador se partem. Pode também ser experimentada sem risco, durante alguns segundos, em aviões em
queda livre segundo trajectórias parabólicas, ou por períodos bem mais longos em veículos orbitando
a Terra.

Muitas actividades humanas habituais à superfície da Terra são incompatíveis com a situação de
imponderabilidade. Um exemplo é a dificuldade em caminhar na Estação Espacial Internacional. Na
Estação os astronautas não são ``puxados" contra o pavimento e, portanto, os seus pés não exercem
qualquer acção (``peso") permanente neste. Em consequência, não há a força normal das paredes e,
portanto, a força de atrito é nula. Como é a força de atrito entre o pavimento e os pés que
proporciona a reacção necessária para andar, não é possível caminhar na Estação.

\subsubsection*{c) \textbf{\emph{a\(>\)g}}}

De acordo com a relação \ref{00}, quando o elevador e a balança se movem com aceleração dirigida
para o centro da Terra de intensidade superior ao valor da aceleração devida à gravidade,
 \(a>g>0\), a força ``peso do corpo" seria:
\begin{equation}\label{33}
     \vec{P}=m(a-g)\hat{z}.
\end{equation}
Isto é, a força ``peso" teria o sentido oposto ao da força da gravidade e o prato da balança seria
puxado no sentido \(+\hat{z}\), oposto à aceleração do elevador, indicando a balança ``peso"
negativo! Esta situação não é fisicamente possível a não ser que o corpo esteja constrangido a
permanecer ligado à balança (porque está colado ou preso a esta). Não se verificando qualquer
ligação forçada o corpo perde o contacto com a balança ficando apenas sujeito à acção da gravidade.
Como consequência, o corpo atrasa-se relativamente à balança, acabando por ser ``apanhado" pelo
tecto do elevador.

Imaginando que existe outra balança no tecto do elevador e que o corpo permanece em contacto com
ela,
a força ``peso do corpo" na balança do tecto é:
\begin{equation}\label{333}
     \vec{P}=m(a-g)\hat{z}, \textrm{~ com ~}a>g.
\end{equation}
A força ``peso do corpo" é agora a acção que o corpo exerce na balança do tecto, e tem a orientação
oposta à força da gravidade exercida pela Terra no corpo, i.e., para o corpo a anterior orientação
``para baixo" passa a ser o sentido ``para cima" e vice-versa.

\section{Força ``peso do corpo" num campo gravitacional fraco}

A acção que um corpo exerce nos seus suportes, a força ``peso do corpo", não depende da existência
de um campo gravitacional na região do espaço onde este se encontra. Considere-se um veículo
espacial numa região do Universo onde o efeito gravitacional é nulo ou pouco significativo, i.e.,
\(\vec{g}\simeq\vec{0}\). Se uma nave em movimento rectilíneo e uniforme nesta região accionar os
seus motores entrará em movimento acelerado e os objectos no seu interior serão ``projectados" no
sentido oposto à aceleração \(\vec{a}\) da nave, tal como acontece quando um carro acelera,
acabando estes por exercer forças nos seus suportes ou nas paredes da nave. Isto é, quando a nave
espacial acciona os motores passa a ser actuada pela força de propulsão que é transferida a cada
objecto no seu interior. Estes, após entrarem em contacto com as paredes da nave ou com outros
corpos solidários com a nave, deslocam-se com a aceleração \(\vec{a}\) da nave. Nestas condições a
equação \ref{00} toma a forma
\begin{equation}\label{0000}
    \vec{P}=-m\vec{a},
\end{equation}
i.e., a acção que o corpo exerce no seu suporte tem o sentido oposto à aceleração da nave e depende
apenas da intensidade dessa aceleração e da massa do corpo.

Pode, portanto, concluir-se que a força que os corpos exercem nos seus suportes ou a ``sua
ausência" não têm a ver necessariamente com o facto do corpo estar ou não sob a influência de um
campo gravitacional. Mesmo na presença de um campo gravitacional significativo, esta força depende
essencialmente das características de movimento do corpo e do seu suporte. De facto, as propostas
para ``criar artificialmente" o efeito da gravidade no interior de naves interplanetárias empregam
a rotação da nave em torno do seu centro de massa, sendo a aceleração devida à ``gravidade
artificial" oposta à aceleração centrípeta da nave.

\section{Imponderabilidade e microgravidade}



Nas situações ideais de queda livre todas as partes de um avião ou de nave espacial acelerariam
uniformemente e o ambiente no seu interior seria de ``gravidade-zero", porque os corpos no seu
interior não sentiriam os efeitos da gravidade. Em situações reais o efeito da gravidade nos
corpos, embora substancialmente reduzido, faz-se sentir, e diz-se que os corpos estão em ambiente
de ``microgravidade" porque a ``ausência de peso" não é total. Contudo, o emprego do termo
``microgravidade" sem especificar o seu significado real pode aumentar os equívocos associados à
identificação da força ``peso" com a força da gravidade.

A total ``ausência de peso" numa nave espacial só seria possível se todos os pontos da nave
estivessem em movimento rectilíneo e uniforme, fora da acção de qualquer campo gravitacional. Neste
caso, os corpos dentro da nave não experimentariam qualquer aceleração uns relativamente aos
outros, e a situação seria de gravidade-zero, desprezando é claro as interacções gravitacionais
mútuas e com as paredes da nave.
As situações de imponderabilidade (``ausência de peso") correntes são frequentemente designadas, de
forma imprópria, como ``gravidade-zero" ou ``microgravidade". É importante ter presente que no
interior de uma nave nas proximidades da Terra a intensidade da aceleração devida à gravidade
terrestre não é nula (``gravidade-zero") ou diminuta (``microgravidade"): a 400 km de altura, por
exemplo, a aceleração devida à gravidade terrestre é 8,4 m s\(^{-2}\), e mesmo à distância da
órbita da Lua é 2,6\(\times10^{-3}\) m s\(^{-2}\), embora nessas regiões a aceleração devida à
gravidade solar seja cerca de duas vezes superior, i.e., aproximadamente 5,8\(\times10^{-3}\) m
s\(^{-2}\). A verdadeira microgravidade terrestre, \(g\approx 10^{-6}\) m s\(^{-2}\), só poderia
ser experimentada em regiões à distância de 17 vezes a separação entre a Terra e a Lua.

Para os cientistas, o termo microgravidade caracteriza o facto de que a nave em queda livre e os
corpos no seu interior não estarem todos sujeitos exactamente à mesma aceleração, em resultado das
diferentes interacções gravitacionais entre os corpos interiores e exteriores à nave, da acção de
outras forças, e traduz-se na aceleração residual que os corpos no interior de um veículo em queda
livre experimentam relativamente, por exemplo, ao centro de massa deste. Em consequência, nas
proximidades da Terra e em naves não tripuladas em queda livre obtém-se com facilidade \(1 \mu g\),
enquanto que em missões tripuladas dificilmente se consegue menos de \(100 \mu g\). As causas mais
relevantes são: i) a variação da aceleração devida à gravidade em resultado da morfologia da Terra;
ii) variações devidas ao efeito gravitacional dos outros corpos celestes, em particular do Sol e da
Lua, dependentes das posições relativamente à Terra
iii) a variação da aceleração da gravidade terrestre com a altitude que decresce aproximadamente 1
parte por milhão por cada 3 m de aumento (numa nave em órbita a força centrípeta e, portanto, a
aceleração devida à gravidade é superior na parte da nave mais afastada da Terra do que na parte
mais próxima);
iv) a atmosfera, embora podendo ser muito rarefeita, por exemplo, a 400 km de altura, desacelera
gradualmente a nave.

\section{Conclusão}

A identificação da força ``peso" com a força da gravidade carece de lógica do ponto de vista do
conhecimento actual e é, muitas vezes, fonte de confusão.
Na definição operacional discutida neste texto, a força ``peso do corpo" corresponde à acção que
este exerce no seu suporte, e depende do estado de movimento de ambos. Tendo presente que o
conceito de ``peso" não é fundamental em Física, acredita-se que a aprendizagem desta ciência sairá
beneficiada se a expressão ``peso do corpo" não for usada. Uma vantagem óbvia é eliminar a confusão
frequente na disciplina entre o conceito de massa do corpo e de força ``peso do corpo".
Espera-se que este texto ajude a motivar a comunidade científica, os professores e os autores dos
manuais de Física, a adoptarem a definição operacional da força ``peso", embora o abandono do
enunciado força ``peso do corpo" seja desejável.



\section*{Agradecimentos}

O autor agradece aos Professores Robertus Potting, Paulo Sá, José Rodrigues, e Paulo Silva os
comentários e a revisão do manuscrito.
